\documentclass[journal,11pt,onecolumn,draftclsnofoot,]{IEEEtran}

\usepackage[T1]{fontenc}

\usepackage{tikz}
\usetikzlibrary{shapes,arrows}
\usepackage{graphicx}
\usepackage{float}
\usepackage{amsmath}
\usepackage[numbers,sort&compress]{natbib}
\usepackage{caption}

\begin{document}

\title{Signal Reconstruction in Diffusion-based Molecular Communication}

\author{Baris~Atakan, Fatih~Gulec ~
	\thanks{B. Atakan and F. Gulec are with the Department of Electrical and Electronics Engineering, Izmir Institute of Technology, Izmir 35430, Turkey (e-mail: barisatakan@iyte.edu.tr; fatihgulec@iyte.edu.tr).} 
	\thanks{\textbf{This paper is submitted to Transactions on Emerging Telecommunications Technologies on January 30, 2019.}}}

\maketitle 



\begin{abstract}
	 Molecular Communication (MC) is an important nanoscale communication paradigm, which is employed for the interconnection of the nanomachines (NMs) to form nanonetworks. A transmitter NM (TN) sends the information symbols by emitting molecules into the transmission medium and a receiver NM (RN) receives the information symbols by sensing the molecule concentration. In this paper, a model of how a RN measures and reconstructs the molecular signal is proposed. The signal around the RN is assumed to be a Gaussian random process instead of deterministic approach in a more realistic way. After the reconstructed signal is derived as a Doubly Stochastic Poisson Process (DSPP), the distortion between the signal around the RN and the reconstructed signal is derived as a new performance parameter in MC systems. The derived distortion which is a function of system parameters such as RN radius, sampling period and the diffusion coefficient of the channel, is shown to be valid by employing random walk simulations. Then, it is shown that the original signal can be satisfactorily reconstructed with a sufficiently low-level of distortion. Finally, optimum RN design parameters, namely RN radius, sampling period and sampling frequency, are derived by minimizing the signal distortion. The simulation results  reveal that there is a trade-off among the RN design parameters which can be jointly set for a desired signal distortion. 
\end{abstract}


\IEEEpeerreviewmaketitle


\section{Introduction}
Molecular Communication (MC) where chemical signals are used instead of  electrical signals to carry information is an emerging area combining biology, chemistry, biophysics and communication engineering. Biologists have been researching in this area for a long time, but the research on MC by communication engineers is still in its infancy. MC can be examined in microscale (nm to cm) and in macroscale (cm to m). Moreover, MC can be performed in aqueous or gaseous environments \cite{Farsad-2016}.

Nanomachines (NMs) are defined as the artificial devices which are composed of the nanometer-scale components. In the existing literature, the term NM mostly refers to bionanomachines, bionanorobots and genetically engineered cells \cite{Atakan-2014}. While a single NM is a very low-end machine with extremely limited capabilities, the interconnection of the NMs forms nanonetworks and makes sophisticated bio/nanotechnology applications possible. It is important for the NMs to communicate among each other to perform more complex tasks \cite{Akyildiz2008, akyildiz2011nanonetworks}. MC is one of the most prominent communication paradigms for the interconnection of the NMs \cite{Hiyama-2005}.


In diffusion-based MC, the transmitter NM (TN) sends information symbols to the receiver NM (RN) by emitting different levels of molecule concentrations. Then, the RN senses the surrounding molecule concentration levels to reconstruct the signal emitted by the TN. In order to understand the performance of the MC systems more clearly, the accuracy of the signal reconstruction needs to be investigated.

In fact, the accuracy of concentration sensing is studied in the biophysics domain where the cell is considered as a molecule concentration measuring device. In the biophysics literature, there are two approaches about how the cell infers information about its environment by sensing the molecule concentration. The first approach is perfect monitoring where the cell is modeled as a permeable sphere and counts the molecules inside its volume. The second approach is  perfect absorbing, where the cell is assumed to count the molecules hitting its surface. The first work about the molecule concentration measurement of a cell is given by Berg and Purcell \cite{Berg-1977}. In this pioneering paper, how a cell can measure the constant molecule concentration as a perfect monitoring device and as a perfect absorbing device with receptors are proposed. The uncertainty, which is defined as the mean square fluctuation of the measured molecule concentration, is derived for a constant molecule concentration outside the cell. In the work of Endres and Wingreen \cite{Endres-2008}, the cell is modeled as a perfect absorber which counts the molecules with the receptors on its surface. Furthermore, the cell is modeled as a gradient sensing device for perfect monitoring and perfect absorbing models. It is stated that the perfect absorbing model is better than the perfect monitoring model both in concentration measurement and gradient sensing, since the previously counted molecules are not counted again and removed from the medium \cite{Endres-2008}. Another method for the concentration measurement is the maximum likelihood estimation (MLE) which is derived by using the probability of time series for the receptor occupancy by the molecules \cite{Endres-2009}. The uncertainty of the estimate is found to be better by a factor of two according to the Berg-Purcell limit given in the work of Berg and Purcell \cite{Berg-1977}. This corresponds to the fact that cells can sense the molecule concentration two times more accurate with MLE. The comparison of these models and sensing limits are given in \cite{Aquino-2016}.



In the existing literature of diffusion-based MC, there are two assumptions about the reception of the RNs. The first one is that the molecule concentration around the RN is assumed to be constant \cite{pierobon2010physical,pierobon2011noise}. In the work of Pierobon and Akyildiz \cite{pierobon2010physical}, the reception process is given as a transformation process. However, there is no derivation of what this process is related to. In another work of Pierobon and Akyildiz \cite{pierobon2011noise}, the molecule concentration outside the RN, which is modeled as a receiver with receptors, is given as a deterministic function and an additive reception noise is defined before the reception which is employed to model the random effects of the molecule-receptor binding process. Similarly, an additive counting noise is defined \cite{pierobon2011diffusion} and employed in the literature \cite{lin2018mutual, mosayebi2018type} to model the error between the constant signal outside the RN and the reconstructed  signal, but the distortion between them is not derived in these studies. The second assumption about the reception of the RNs, is that the molecule concentration is assumed to be sensed perfectly by the RNs \cite{kilinc2013receiver, srinivas2012molecular, noel2014optimal, mustam2017multilayer}. However, these two assumptions cannot be realistic due to the stochastic nature of the molecule movements. None of the studies in the diffusion-based MC literature assumes the molecule concentration as a random process and derives an error, which can occur during the molecule sensing process of the RN. In this study,  a novel approach is proposed without any need for these assumptions. The existing molecule concentration around the RN is modeled as a Gaussian random process resulting the reconstructed signal as a Doubly Stochastic Poisson Process (DSPP). Although the DSPP is used to model the input signal in neuro-spike communications \cite{aghababaiyan2018axonal, maham2018neuro}, the random processes are not adopted to model the signal outside the RN in diffusion-based MC. With the random process approach, the signal reconstruction performance is investigated by deriving a signal distortion function, which is the Mean Square Error (MSE) between the existing signal around the RN and the reconstructed signal. By using the random walk simulation of the molecules, the derived distortion function, which consist of the system parameters such as the RN radius, diffusion coefficient and sampling period, is validated. In addition, the distributions of the original and the reconstructed signals are generated. The results show that the RN can reconstruct the surrounding signal with a small distortion, if the system parameters are appropriately selected. Besides the derivation of the distortion  during the signal reconstruction, our work contributes to the literature by revealing the relation between the signal reconstruction and the RN design. We obtain optimum RN design parameters using the cases where the signal distortion function is minimum with respect to the corresponding parameter. The trade-off among the RN design parameters is shown with the numerical results. Through the extensive analytical and numerical analyses, the optimal design parameters of the RN such as the RN radius and sampling frequency are investigated by minimizing the signal distortion. 


The remainder of the paper is organized as follows. In Section \ref{Motivation}, the motivation to find the accuracy of the molecular signal reconstruction is given. In Section \ref{System Model}, the system model for the signal reconstruction is introduced. The distortion of the reconstructed signal in the MSE sense is derived in Section \ref{Derivation}. The validation of the system model and  the numerical results are presented in Section \ref{Simulation}. In Section \ref{RN Design}, the optimum design parameters of the RN  are derived and analyzed. Finally, the concluding remarks are given in Section \ref{Conclusion}.
\section{Motivation}
\label{Motivation}
In this section, our motivation for the system model is explained by using a one dimensional scenario. Let us consider a TN emitting molecules instantaneously as a single spike to send an information symbol to the RN. The molecule concentration around RN can be given by
\begin{equation}
	C = \dfrac{Q}{\sqrt{4 \pi D t}} e^{\frac{-r^2}{4Dt}},
	\label{C}
\end{equation}
where $ Q $ is the number of the molecules, $C$ is the molecule concentration, $ D $ is the diffusion coefficient, $ r $ is the distance from the TN and $t$ is the propagation time \cite{bossert1963analysis}.

\begin{figure}[!hbt]
	\centering
	\scalebox{0.45}{\includegraphics{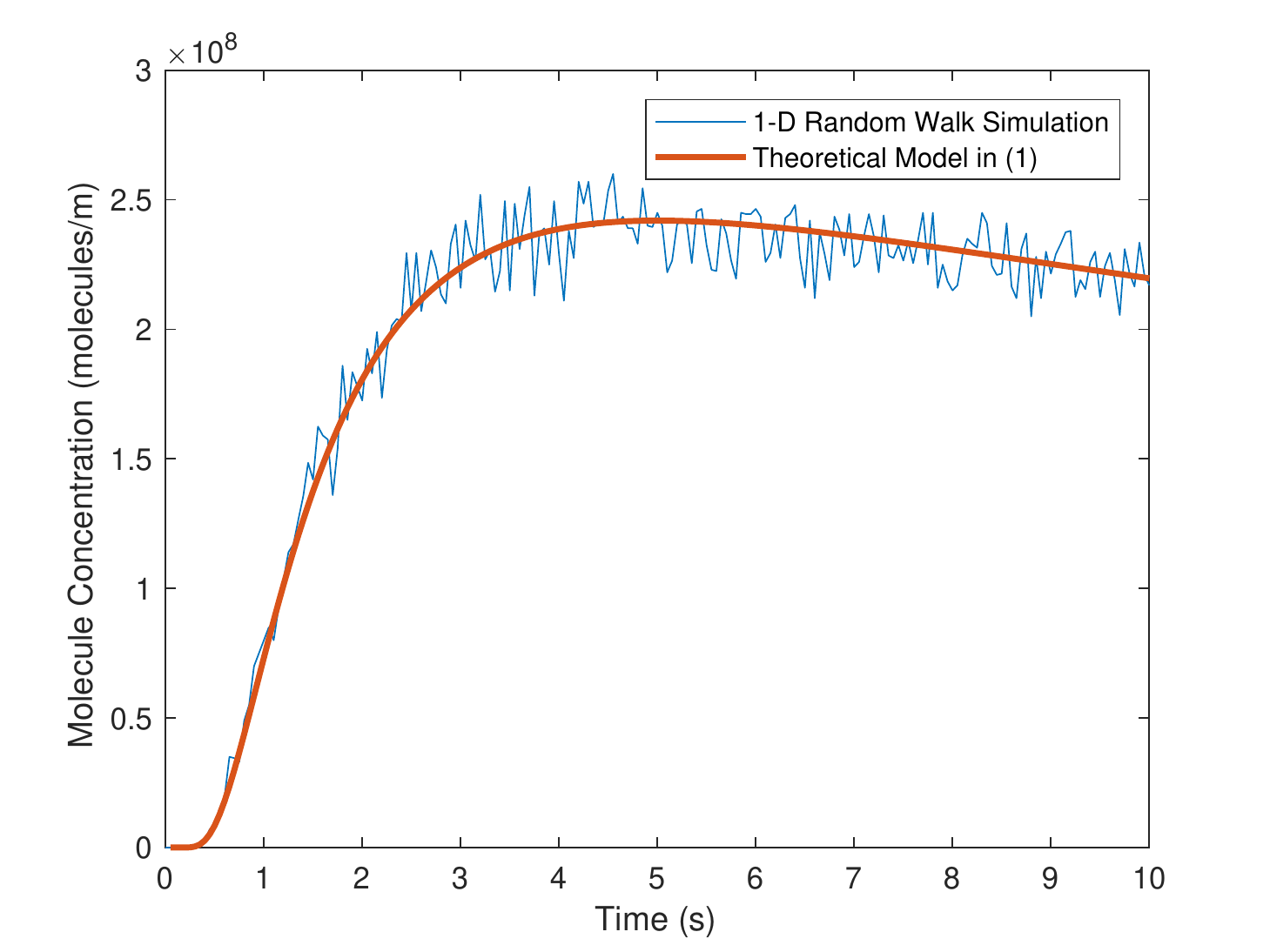}}
	\caption{The comparison of the theoretical model and the random walk simulation for the molecule concentration with the parameters $Q=10^4$, $ D=10^{-11}m^2/s $, $r=10 \mu m$, (step time) $\tau = 10^{-3} s$, (step length) $\delta = 0.0447 \mu m$.}
	\label{CR}
\end{figure}

Actually, (\ref{C}) characterizes the average behavior of molecule concentration at a certain distance from the TN. However, the instantaneous molecule concentration changes randomly due to the random walk phenomenon which is utilized to model the diffusion of the molecules. According to this phenomenon, molecules make random steps to the right or to the left direction with equal probability on the $x$ axis. Their successive steps are statistically independent from their previous steps.
The random movements cause the molecule concentration to change randomly at each instant of the diffusion. Therefore, the molecule concentration generates a random process rather than a constant value given in (\ref{C}). This is illustrated in Fig. \ref{CR} where the solid line represents the theoretical model in (\ref{C}) and the oscillating line represents the random process generated via a random walk simulation. Due to the randomness of the molecules, the molecule concentration needs to be modeled by a random process instead of a deterministic function. Therefore, our main motivation is to investigate how accurate the RN reconstructs the surrounding signal, which is modeled as a random process, as given in the next section.
\section{System Model}\label{System Model}
\begin{figure}[!t]
	\centering
	\scalebox{0.45}{\includegraphics{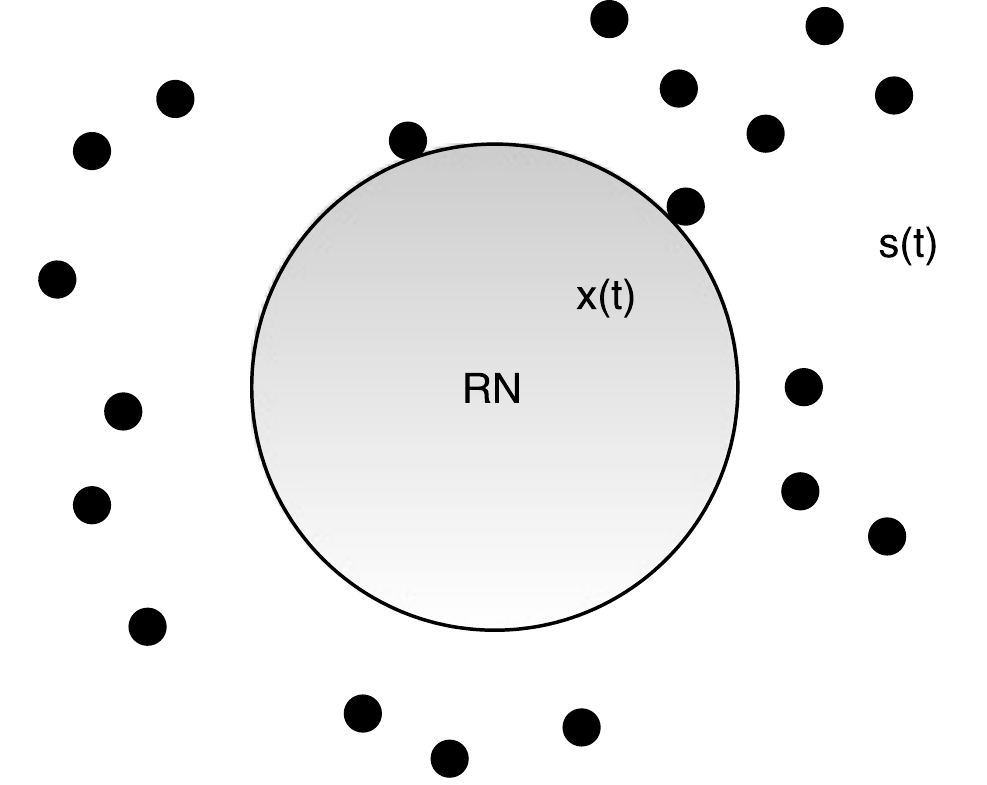}}
	\caption{Signal reconstruction model for the RN.}
	\label{NM}
\end{figure}

\begin{figure}[!t]
	\centering
	\scalebox{0.45}{\includegraphics{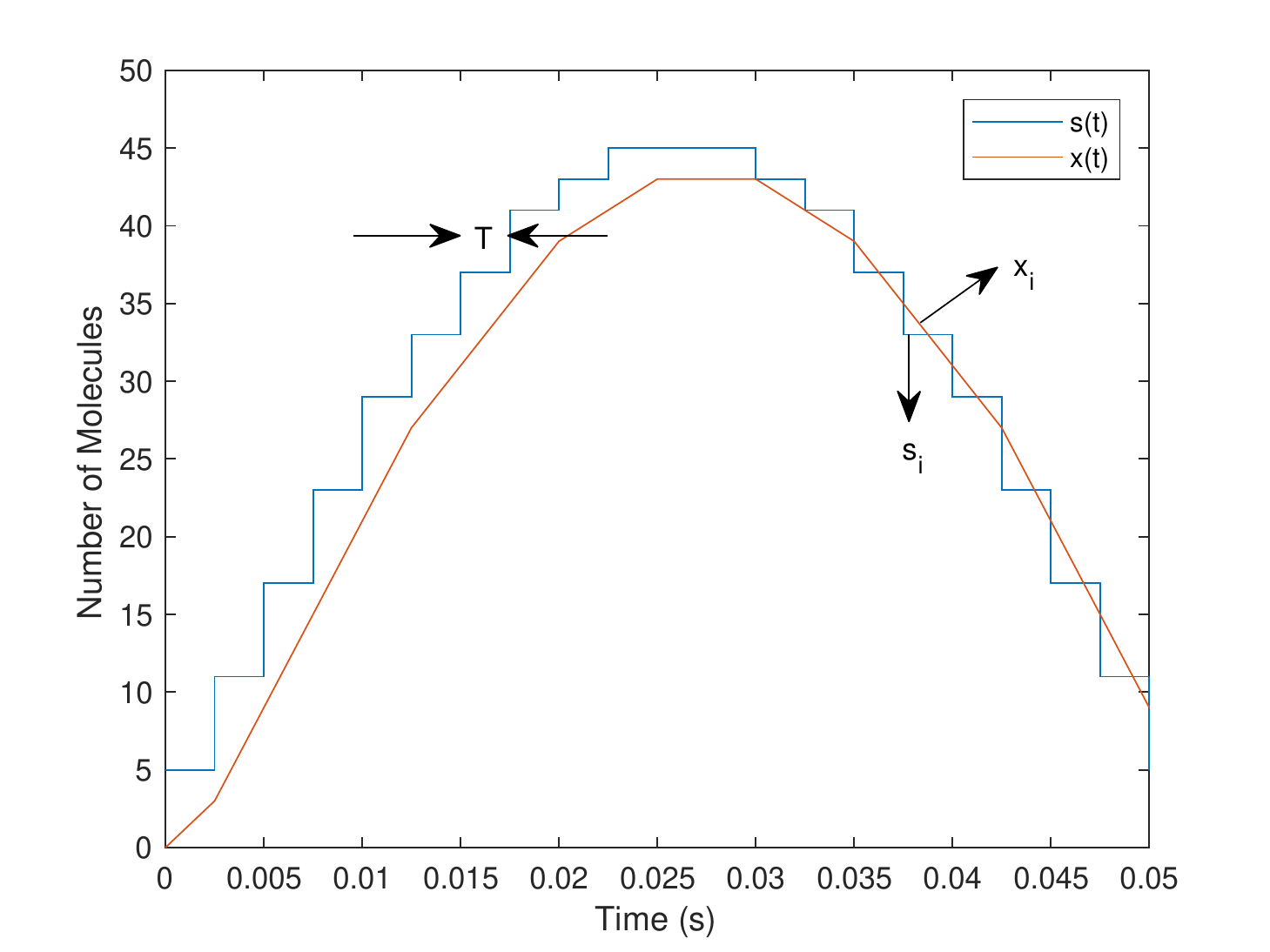}}
	\caption{Existing signal outside the RN and the sampled signal.}
	\label{Delta}
\end{figure}

A model to perceive and reconstruct the signal around the RN is proposed in this section. The RN is assumed to be a perfect absorber meaning that a molecule is received, when it hits its surface. In this way, the RN senses the molecule concentration during certain observation periods, i.e., samples the signal. It is also assumed that no chemical reaction occurs during the movement of the molecules. Let $s(t)$ be the random process that shows the number of the molecules outside the RN and $x(t)$  be the counted number of the molecules by the RN. This model is illustrated in Fig. \ref{NM}. $ s(t) $ is assumed as a  Poisson process. Due to the large number of the molecules, the Poisson distribution can be approximated by the Gaussian distribution. Hence, $ s(t) $ is assumed to be a Gaussian random process with mean $m(t)$, variance $v(t)$ and autocovariance $R(t_1, t_2)$. The RN counts the molecules along a sampling period $T$ and it is assumed that the number of the molecules does not change in this period. An analogy with the delta modulation can be established to understand the relation between $s(t)$ and $x(t)$. As illustrated in Fig. \ref{Delta}, $s_i$ and $x_i$, which are measured along $ T $, are the $i^{th}$ samples of $s(t)$ and $x(t)$, respectively. Since $s(t)$ is assumed to be a Gaussian random process, $s_i$ is assumed as a Gaussian random variable with $N(\mu_s, \sigma_s^2)$.
\begin{figure}[!b]
	\centering
	\scalebox{0.6}{\includegraphics{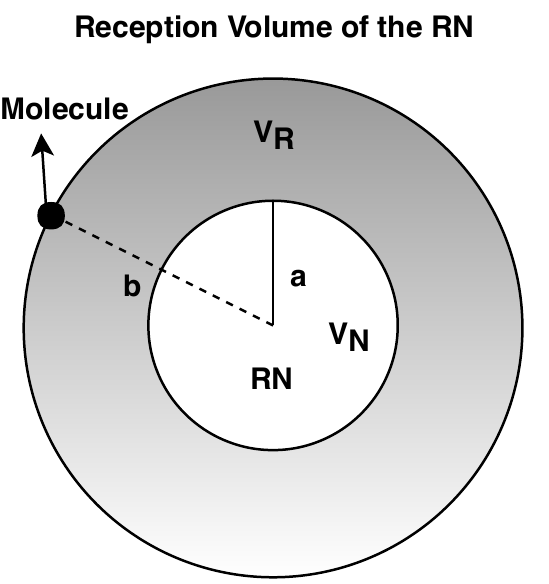}}
	\caption{Reception volume of the RN and a single molecule.}
	\label{RV}
\end{figure}

In order to define a molecular signal as the changing concentration levels around the RN, it is essential to specify a volume in which the RN is located similar to the virtual reception volume approach employed in \cite{atakan2010deterministic}  as illustrated in Fig. \ref{RV}. In this figure, $V_N$ denotes the volume of the RN and $V_R$ stands for the reception volume, in which the molecular signal exists. $a$ and $b$ are the radii of the spherical volumes $V_N$ and $V_R$, respectively. When a molecule is received by the RN, it is not released to $V_R$ again. The probability that the molecule emitted at the distance $y$ from the center of the RN can reach its surface within time $t$ can be found through the first-hitting time probability of the random walk as follows \cite{ziff2009capture, bai2016throughput, yilmaz2014Threed},
\begin{equation}
	F(y,t) = \dfrac{a}{y} \textrm{erfc} \bigg(\dfrac{y-a}{\sqrt{4Dt}}\bigg),
	\label{CDF}
\end{equation}
where $D$ is the diffusion coefficient, $ a $ is the radius of the RN and erfc(.) is the complementary error function. This capture can be used to find the hitting rate of the molecules to the surface of the RN. Since the capture of a molecule is an event with two possible outcomes as "capture" or "escape", $x_i$ can be assumed as a random variable with a binomial distribution. Due to the large number of molecules around the RN, the binomial distribution can be approximated to Poisson distribution. Thus, $x_i$ is assumed to have a Poisson distribution in our study. On the other hand, the spatial distribution of the molecules around the RN is given as follows. It is assumed that the molecules in the reception volume are uniformly distributed over the interval $ (a, b) $ where $b > a$. By using the mean distance of the molecules to the surface of the RN, i.e., $\frac{a+b}{2}$, the average number of molecules which can hit the surface of the RN within a sampling interval can be given as,
\begin{equation}
	\lambda = F\left(\frac{a+b}{2}, T\right)s_i,
	\label{Rate}
\end{equation}
where $ F(\frac{a+b}{2}, T)$  is abbreviated as $ F(T) $ in the rest of the paper. Here, $\lambda$ is the rate of the Poisson random variable $x_i$ depending on another random variable $s_i$.

A Poisson process is generally employed as a counting process and it can be defined as a random process which has independent increments being Poisson distributed. The rate of the Poisson process can be generalized as the time-varying intensity function, i.e., $\lambda(t)$. When the intensity function is constant, the Poisson process is homogeneous, whereas the Poisson process is nonhomogeneous when the intensity function is time-dependent. The intensity function may be unknown in some situations and cannot be treated as constant. In such circumstances, it is reasonable to consider the intensity function as a random process. When the intensity function of the nonhomogeneous Poisson process is another random process, the Poisson process is called a Cox process or a Doubly Stochastic Poisson Process (DSPP) \cite{cox1955some}.

In our system model, the intensity function of $ x(t) $  can be found  by replacing the Gaussian random variable $ s_i $   with the Gaussian random process $ s(t) $ in (3). This intensity function is given by
\begin{equation}
	\lambda(t) = \frac{F(T)}{T}s(t),
	\label{intensity}
\end{equation}
where the intensity function is normalized by a factor of $ \frac{1}{T} $ due to the definition of the rate, which is the number of molecules hitting the surface of the RN within unit time. Since the intensity of $x(t)$ is another random process given in (\ref{intensity}), $x(t)$ is a DSPP. Furthermore, $s(t)$ can be assumed as a stationary random process, since it is constant along a sampling period $T$.

Using the knowledge that $x(t)$ is a DSPP which is defined for $t \geq t_0$ and $s(t)$ is a stationary Gaussian random process, the mean of $x(t)$ can be given as \cite{Valderrama-1995} 
\begin{align}
	E\{x(t)\} &= \int_{t_0}^{t}E\{\lambda(u)\} du = \int_{t_0}^{t}E\left\{\frac{F(T)}{T}s(u)\right\} du \notag\\
	&= \frac{F(T)}{T} \int_{t_0}^{t}m(u) du,
	\label{Mean_x}
\end{align}
where $E\{.\}$ represents the expectation operator. To find the variance of $x(t)$, the autocovariance of $\lambda(t)$ is needed which is derived as
\begin{align}
	Cov\{\lambda(t_1),\lambda(t_2)\} &= E\left\{\left[\lambda(t_1) - E\{\lambda(t_1)\} \right] \left[\lambda(t_2) - E\{\lambda(t_2)\} \right]\right\} \notag\\
	&= E\left\{\frac{F(T)}{T} \left[ s(t_1) - m(t_1) \right] \frac{F(T)}{T} \left[s(t_2) - m(t_2) \right]\right\} \notag\\
	&= \frac{\left(F(T)\right)^2}{T^2} R(t_1,t_2),
	\label{cov}
\end{align}
where $ Cov\{.\}$ shows the covariance operator. Then, by using (\ref{cov}), the variance of $x(t)$ is given as \cite{Valderrama-1995} 
\begin{align}
	Var\{x(t)\} &= 2\int_{t_0}^{t} \int_{t_0}^{t_2}Cov\{\lambda(t_1),\lambda(t_2)\} dt_1 dt_2 +\int_{t_0}^{t}E\{\lambda(u)\} du \notag\\
	&= \frac{2\left(F(T)\right)^2}{T^2} \int_{t_0}^{t} \int_{t_0}^{t_2}R(t_1,t_2) dt_1 dt_2 +  \frac{F(T)}{T} \int_{t_0}^{t}m(u) du,
	\label{Variance_x}
\end{align}
where $Var\{.\}$ is the variance operator. The second moment of $x(t)$ can be calculated by using (\ref{Mean_x}) and (\ref{Variance_x}) in the formula $E\{x^2(t)\} = Var\{x(t)\} + \left( E\{x(t)\}\right)^2$ as given by
\begin{equation}
	E\{x^2(t)\} = \frac{2\left(F(T)\right)^2}{T^2}  \int_{t_0}^{t} \int_{t_0}^{t_2}R(t_1,t_2) dt_1 dt_2 +  \frac{F(T)}{T} \int_{t_0}^{t}m(u) du + \left( \frac{F(T)}{T} \int_{t_0}^{t}m(u) du \right)^2.
	\label{x2}
\end{equation}

The statistical properties of the DSPP $x(t)$ is employed to derive the signal reconstruction distortion in the next section.

\section{Derivation of the Signal Distortion}
\label{Derivation}
In this section, the signal distortion between the signal outside the RN and the reconstructed signal is derived as a Mean Square Error (MSE). The signal distortion, i.e., the MSE ($\mathcal{E}$), is given by 
\begin{equation}
	\mathcal{E} =  \dfrac{E\{(s_i - x_i)^2\}}{V^2} = \dfrac{E\{s_i^2\} - 2E\{s_i x_i\} + E\{x_i^2\}}{V^2},
	\label{MSE}
\end{equation}
where $s_i$ is a Gaussian random variable with $N(\mu_s, \sigma_s^2)$ showing the number of the molecules in the volume between the outer boundary of the reception volume and the RN, $(V_R-V_N)$, $x_i$ is a doubly stochastic Poisson variable showing the counted number of the molecules in $V_N$, $V$ is the volume. Here, $  \mathcal{E} $ is derived by using concentrations of the molecules inside and outside the RN. Therefore, (\ref{MSE}) includes the volume $V$ where those molecules are located. Also in (\ref{MSE}), $V_R$ is calculated as $\frac{4}{3}\pi b^3$ and similarly $V_N$ is calculated as $\frac{4}{3}\pi a^3$. Since the counted number of the molecules by the RN  depends on the number of the molecules outside the RN, random variables $x_i$ and $s_i$ are dependent. The second term of the numerator in (\ref{MSE}) consists of the product of these two dependent random variables. The expected value of this product can be found by
\begin{equation}
	E\{s_i x_i\} = \rho_{s_i x_i} \sigma_{x_i} \sigma_{s_i} + E\{x_i\}E\{s_i\},
	\label{sx}
\end{equation}
where $ \rho_{s_i x_i} $ is the correlation coefficient between $x_i$ and $s_i$. After substituting (\ref{sx}) in (\ref{MSE}), the volumes can be clarified to find the molecule concentrations by writing explicitly the volumes for each term as given by
\begin{equation}
	\mathcal{E} = \frac{E\{s_i^2\}}{(V_R-V_N)^2} - 2\rho_{s_i x_i} \frac{\sigma_{x_i}}{V_N} \frac{\sigma_{s_i}}{(V_R-V_N)} - 2 \frac{E\{x_i\}}{V_N} \frac{E\{s_i\}}{(V_R-V_N)} + \frac{E\{x_i^2\}}{V_N^2}.	
	\label{MSE_sx}
\end{equation}

In (\ref{MSE_sx}), the numerator of the first term can be found as $E\{s_i^2\} = \sigma_s^2 + \mu_s^2$. To find the second and the third term of the numerator in (\ref{MSE}), the first and second moments of $x(t)$ is employed. $ E\{x_i^2\} $ can be found as given in equation (\ref{2nd-moment of x_i}) by setting $t_0=iT$ and $t=(i+1)T$ in (\ref{x2}) where $i\geq 0$ is an integer.  Subsequently, the signal distortion becomes as given by (\ref{MSE_long}).
\begin{multline}
	E\{x_i^2\} = E\{x^2(t)\}\bigg|_{t_0=iT}^{t=(i+1)T} =  \frac{2 \left(F(T)\right)^2}{T^2} \int_{iT}^{(i+1)T} \int_{iT}^{t_2}R(t_1,t_2) dt_1 dt_2 \\ + \frac{F\left(T\right)}{T} \int_{iT}^{(i+1)T}m(u) du +  \left( \frac{F\left(T\right)}{T} \int_{iT}^{(i+1)T}m(u) du \right)^2.
	\label{2nd-moment of x_i}
\end{multline}

\begin{figure*}[!b]
	\hrulefill
	\begin{multline} 
		\mathcal{E} = \frac{\sigma_s^2 + \mu_s^2}{(V_R-V_N)^2} - 2 \Bigg[ \frac{\mu_s F\left(T\right)}{(V_R-V_N)T}  \int_{iT}^{(i+1)T}\frac{m(u)}{V_N} du  \\ + \frac{\rho_{s_i x_i} \sigma_s}{(V_R-V_N)V_N}\! \sqrt{\!\int_{iT}^{(i+1)T}\!\int_{iT}^{t_2}\!\frac{2 \left(F(T)\right)^2R(t_1,t_2)}{T^2} dt_1 dt_2\!+\!  \int_{iT}^{(i+1)T}\frac{F\left(T\right)m(u)}{T} du} \Bigg] \\+ \frac{1}{V_N^2} \Bigg[ \frac{2 \left(F(T)\right)^2}{T^2}  \int_{iT}^{(i+1)T} \int_{iT}^{t_2}R(t_1,t_2) dt_1 dt_2 + \frac{F\left(T\right)}{T}\int_{iT}^{(i+1)T}m(u) du  +  \bigg( \frac{F\left(T\right)}{T}\int_{iT}^{(i+1)T}m(u) du \bigg)^2 \Bigg]. 
		\label{MSE_long}
	\end{multline}
	
	\vspace*{4pt}
\end{figure*}

Since $ s(t) $ is assumed to be a stationary random process, the mean, variance and autocovariance can be written as $ m(t) = \mu_s $ and $ v(t) = R(t_1,t_2) = \sigma_s^2 $ for one sample measured along $ T $. Using this assumption, $ \mathcal{E} $ can be simplified in a time interval from $ 0 $ to $ T $ as given in (\ref{MSE_last}) which is used to obtain the numerical results in the next section.

\begin{figure*}[!t]
	\normalsize
	
	\vspace*{4pt}
	
	\begin{equation}
		\label{MSE_last}
		\mathcal{E} = \frac{\sigma_s^2 + \mu_s^2}{(V_R-V_N)^2} - \frac{2 \rho_{s_i x_i} \sigma_s \sqrt{F\left(T\right) \bigg[F\left(T\right)\sigma_s^2 + \mu_s\bigg]} - 2 F(T) \mu_s^2}{V_N (V_R-V_N)} +  \frac{F\left(T\right) \bigg[F\left(T\right)\sigma_s^2 + \mu_s + F\left(T\right)\mu_s^2 \bigg]}{V_N^2}.
	\end{equation}
	\hrulefill
	
\end{figure*}

%
%

\section{Validation of the Signal Distortion Function \& Numerical Results}
\label{Simulation}

In this section, the derived $ \mathcal{E} $ given in (\ref{MSE_last}) is validated through random walk simulations. In addition, the numerical results are given and analyzed by evaluating the signal distortion function. Furthermore, the distributions of the signal outside the RN and the reconstructed signal are observed. The simulation parameters for the numerical results are given in Table \ref{Table-Parameters} and the random walk simulation parameters are given in Table \ref{Table-Parameters_RW}. In the simulation experiments, the signal outside the RN is calculated by dividing the number of molecules $s_i$ to the volume between the boundary of $V_R$ and the boundary of $ V_N $ as shown in Fig. \ref{RV}. Similarly, the concentration received by the RN is found by dividing the received number of molecules $x_i$ to $ V_N $.
\begin{table}[!htb]
	\centering
	\caption{Simulation Parameters.}
	\vspace{2mm}
	\label{Table-Parameters}
	{\renewcommand{\arraystretch}{1.2}
		\renewcommand{\tabcolsep}{0.5cm}
		\begin{tabular}{p{145pt}|p{55pt}} \hline \hline
			\textbf{Parameters}                        & \textbf{Values}    \\ \hline \hline
			$ \mu_s $					          & $ 10^{2} $         \\ \hline
			$ \sigma_s^2   $                               & $ 10^{2} $    \\ \hline
			Correlation coefficient ($ \rho_{sx} $) & 0 - 1    \\ \hline
			Diffusion coefficient ($ D $)				& $10^{-12} m^2/s$\\ \hline
			Radius of the $ V_N $ ($ a $)					& $1 \mu m$ \\ \hline
			Radius of the $ V_R $ ($ b $)					& $2-3 \mu m$ \\ \hline
			Sampling period ($ T $)                   	  & 0 - 0.25 s \\ \hline \hline
	\end{tabular}}
\end{table} 

\begin{table}[!hbt]
	\centering
	\caption{Random Walk Simulation Parameters.}
	\vspace{2mm}
	\label{Table-Parameters_RW}
	{\renewcommand{\arraystretch}{1.2}
		\renewcommand{\tabcolsep}{0.5cm}
		\begin{tabular}{p{145pt}|p{55pt}} \hline \hline
			\textbf{Parameters}                        & \textbf{Values} \\ \hline \hline
			$ \mu_s $					          & $ 10^{2} $         \\ \hline
			Diffusion coefficient ($ D $)				& $10^{-12} m^2/s$\\ \hline
			Radius of the $V_N$ ($ a $)					& $1 \mu m$ \\ \hline
			Radius of the $V_R$ ($ b $)					& $2-3 \mu m$ \\ \hline
			Time ($ t $)                   	  & 0 - 0.25 s\\ \hline
			Step time ($\tau$)						& $10^{-3}$ s \\ \hline
			Step length ($\delta$)					& $0.0447 \mu m$\\ \hline \hline
	\end{tabular}}
\end{table} 

The aim of the numerical result part is to observe the accuracy of the signal reconstruction of the RN for different MC system parameters, such as radius of the reception volume, sampling period and diffusion  coefficient. In the light of these results, the signal reconstruction performance can be improved by adjusting the MC system parameters appropriately. 




\subsection{Validation of the Theoretical System Model}

A random walk simulation is performed to show that the signal distortion function is valid. In the random walk simulations, a molecule is assumed to make a random movement  in every $\tau$ seconds with a step length of $ \delta $ meters on  $ x $, $ y $ and $ z $ axes separately. Every step of a molecule, whose step length can be calculated by $\delta = \sqrt{2D\tau}$, is independent from its other steps \cite{berg1993random}.

\begin{figure*}[!t]
	\centering
	\includegraphics[width=0.375\textwidth]{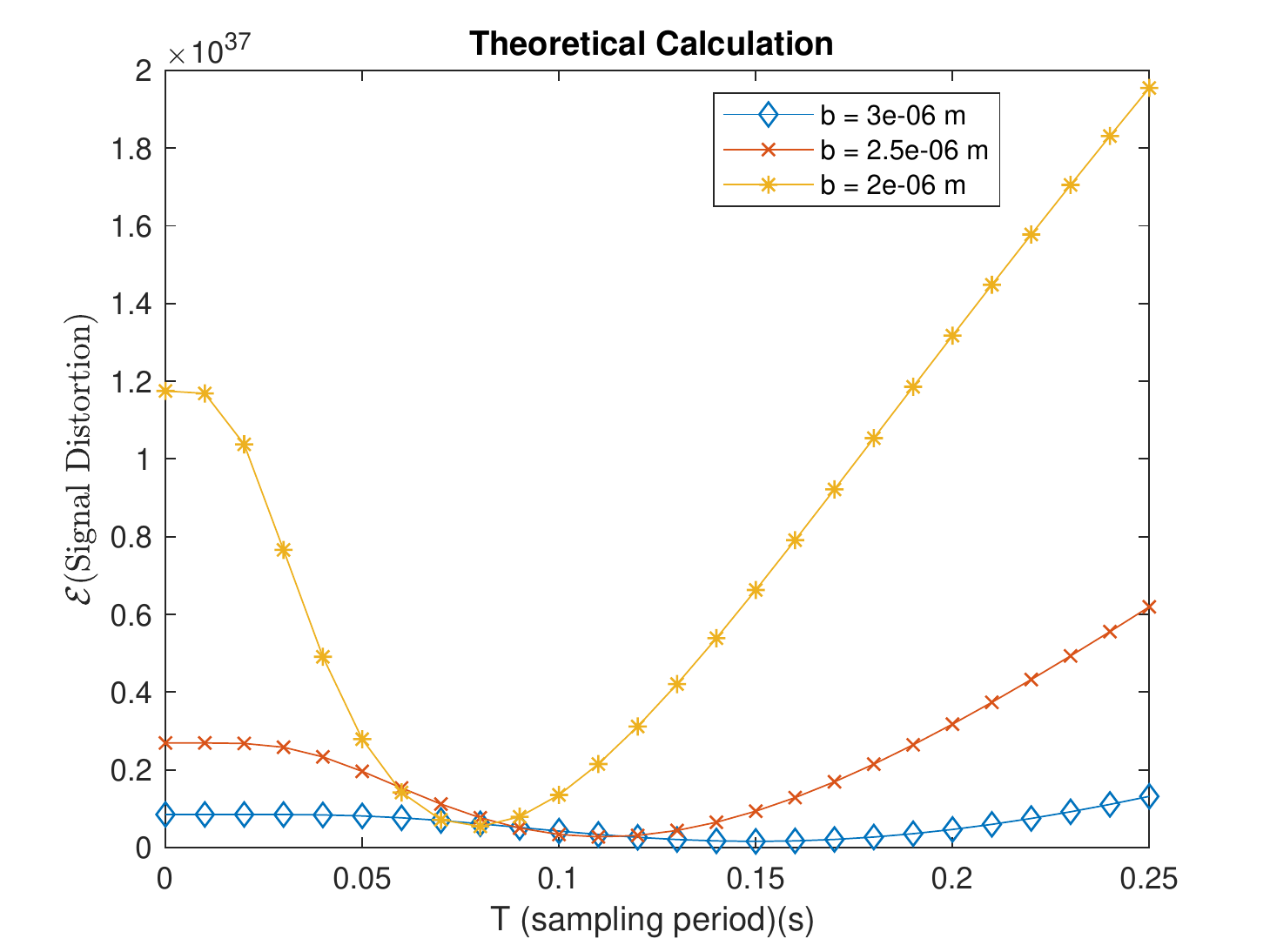}    
	\includegraphics[width=0.375\textwidth]{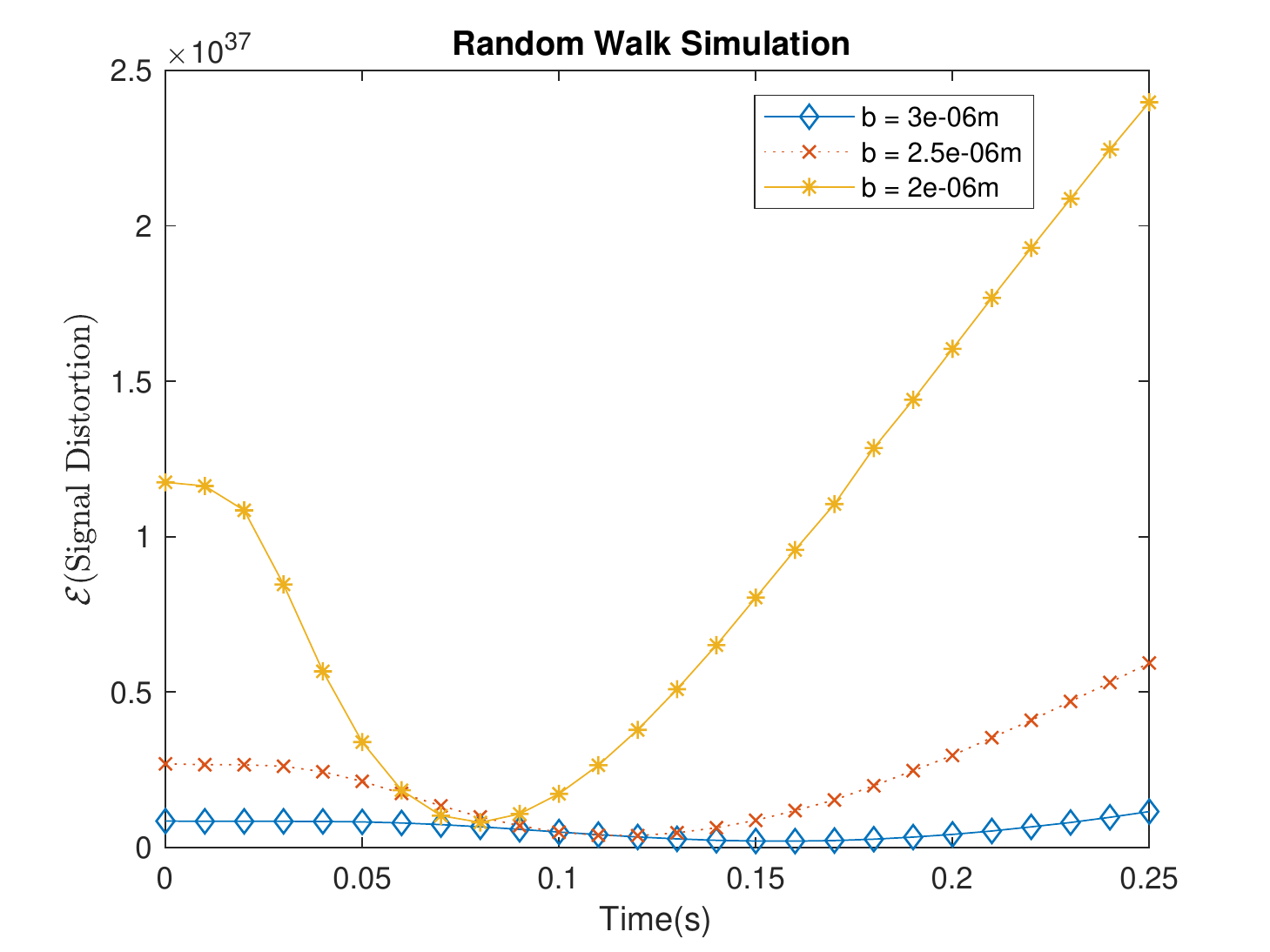} \\
	\scriptsize \hspace{0.1 in}  (a) \hspace{2.3 in} (b)
	\caption{Signal distortion vs. sampling period for different $V_R$ radii (a) theoretical calculation with $  \rho_{sx} = 0.75 $ (b) random walk simulation.}
	\label{MSE_b}
\end{figure*}

During the random walk simulation, the following assumptions are made. The initial distance of the molecules  to the center of the RN is assumed as $(a+b)/2$ which is the midpoint of the boundaries of $V_R$ and $V_N$, as given in the theoretical model. A molecule in $V_R$ is received, when it hits the RN. After the reception of a molecule, it is not released to $V_R$ again. Every received molecule is only counted once, since the RN is assumed to be a perfect absorber. Until the end of the random walk simulation, no additional molecule can enter the $V_R$. Using these assumptions, the numerical and the random walk simulation results for the signal distortion ($ \mathcal{E} $) as a function of the sampling period are given in Fig. \ref{MSE_b} (a) and (b), respectively. In both of the figures, the signal distortion functions have  approximately the same convex structure and nearly the same minimum points. This shows that the signal distortion function given in (\ref{MSE_last}) is valid. Small differences are observed in signal distortion values between Fig. \ref{MSE_b} (a) and Fig. \ref{MSE_b} (b) which can be clarified as follows. In the random walk simulation, the molecules start walking from the midpoint between the boundaries of $V_R$ and $V_N$ and during the sampling period they can move out of the $V_R$. However, in the theoretical model, it is assumed that during the sampling period the number of the molecules is constant and the number of the molecules inside the $V_R$ can decrease if and only if the molecules are received by the RN.

The numerical results given in this section validates the signal distortion function derived in Section \ref{Derivation}  for the system model given in Section \ref{System Model}. Using this signal distortion function, the effect of the system parameters is examined in the next subsection.
\subsection{Numerical Results}
In this part, we first observe the signal distortion function for three different $V_R$ radii as given in Fig. \ref{MSE_b} (a). The different molecular signals (concentration values) are obtained by changing the radius of $V_R$, i.e., $b$, while keeping the number of molecules in $V_R$ constant. As the $V_R$ becomes smaller, the signal distortion values fall more steeply to their minimum values, since the hitting rate of the molecules increases in a smaller volume. This also shows that the RN can respond faster for smaller values of $V_R$. As the RN samples for a longer duration, it can capture more molecules. However, after reaching the minimum of the signal distortion function  which  gives the optimum sampling period on the x-axis, the RN captures more molecules than it needs to calculate the valid molecule concentration around it. Therefore, the signal distortion increases after the minimum of the signal distortion function. This result reveals that the RN can increase its signal reconstruction performance by increasing the sampling period, only until the minimum signal distortion point.

\begin{figure*}[!t]
	\centering
	\includegraphics[width=0.375\textwidth]{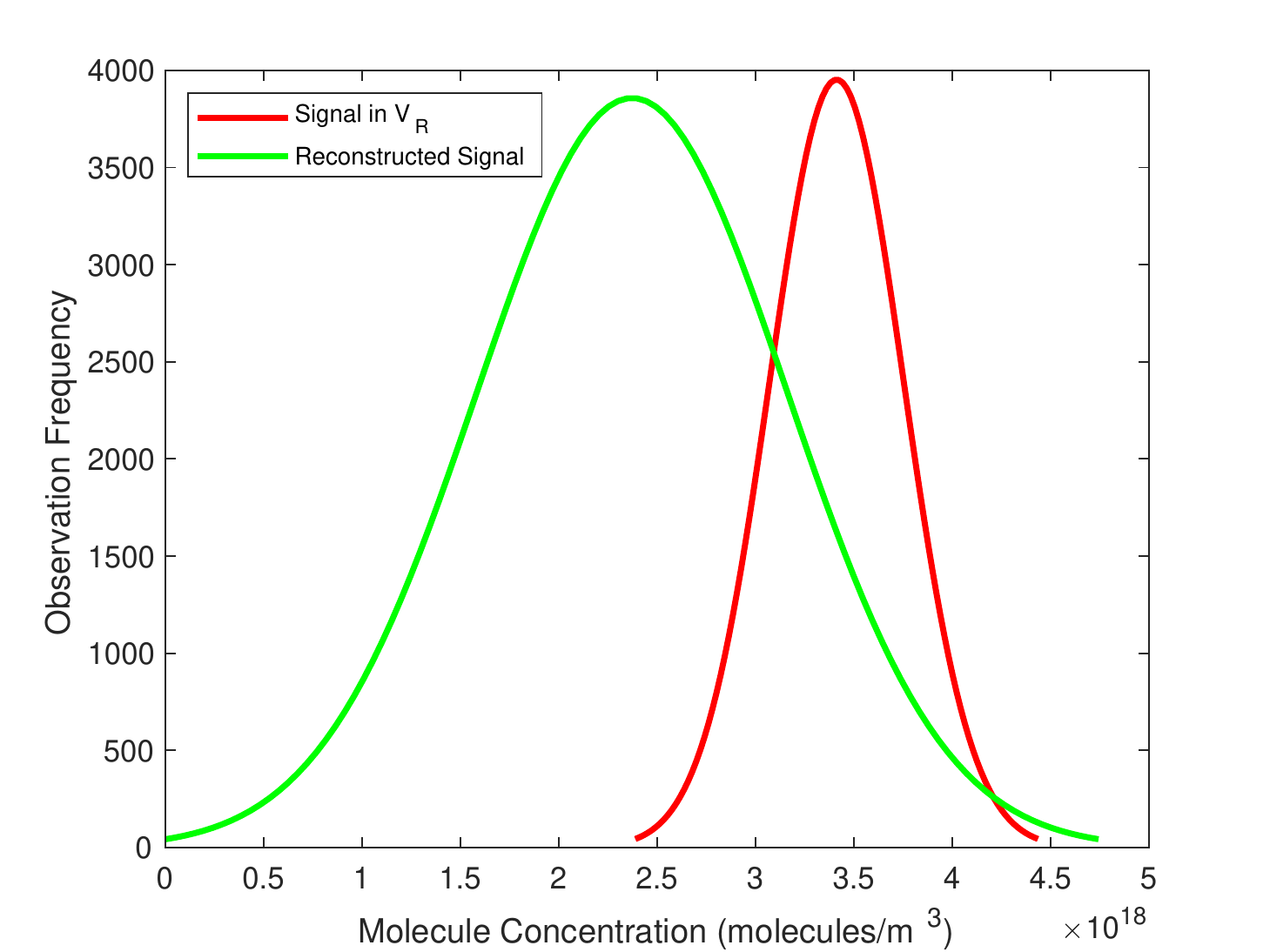}    
	\includegraphics[width=0.375\textwidth]{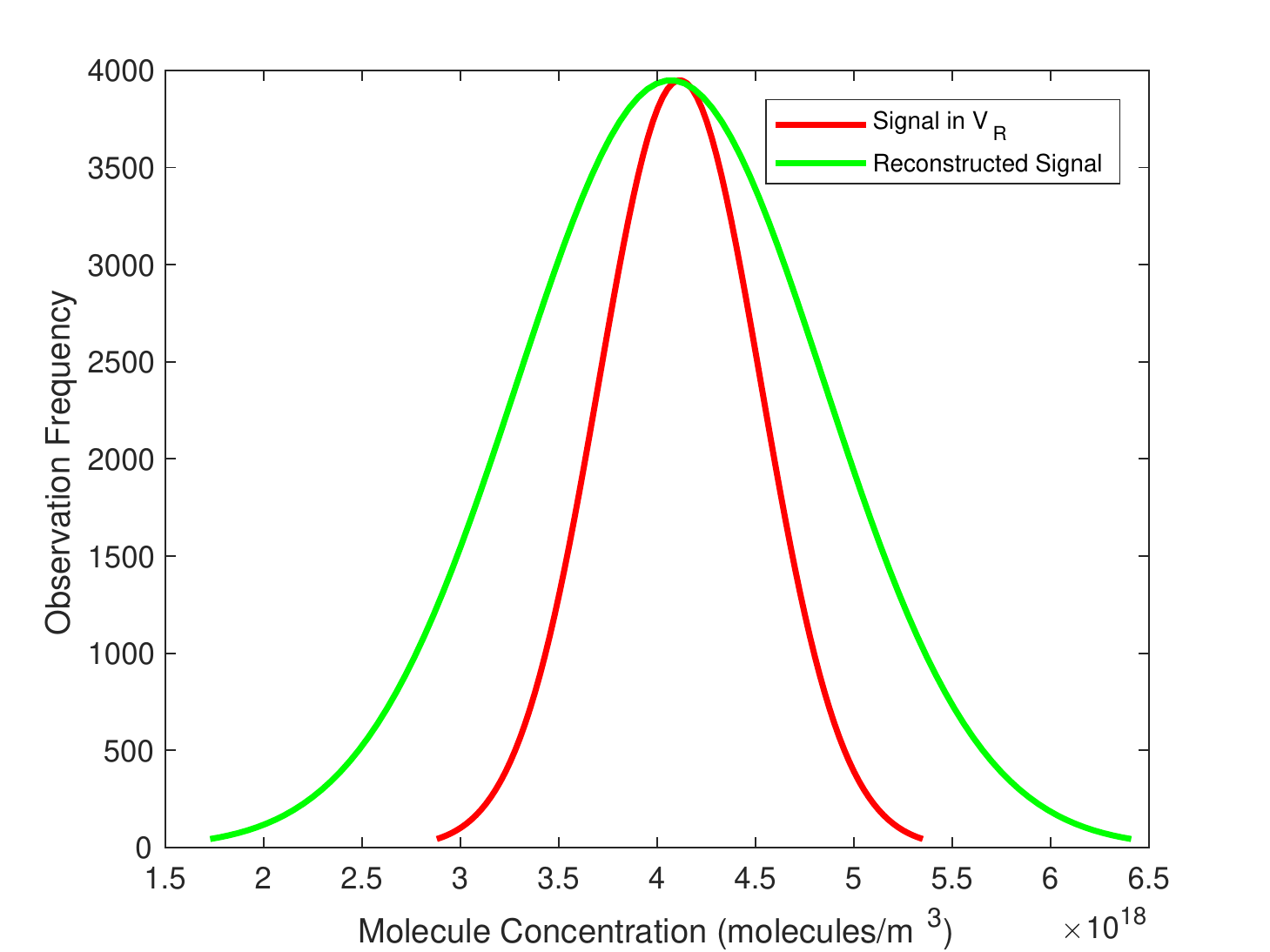} \\
	\scriptsize \hspace{0.1 in}  (a) \hspace{2.3 in} (b)
	\caption{Distributions of $ x $ and $s $ with the parameters (a) $ T=0.06 s $, $D = 10^{-12} m^2/s$,	$ \mu_s = 10^2$, $\sigma_s^2 = 10^2 $, $ a=1 \mu m $ , $b=2 \mu m$ (b) $ T=0.12 s $, $D = 10^{-12} m^2/s$,	$ \mu_s = 10^2$, $\sigma_s^2 = 10^2 $, $ a=1.3 \mu m $ , $b=2 \mu m$.}
	\label{Distsx}
\end{figure*}

In Fig. \ref{Distsx} (a) and (b), the fitted distributions of the original signal samples, $ s $, and the reconstructed signal samples, $ x $, are shown for different parameters. The results are obtained as follows. First, a Gaussian distribution for $s$ is generated. Then, by using the distribution of $s$, the distribution of $x$ is calculated by generating a Poisson distribution with the rate given in (\ref{Rate}) for $10^6$ trials. The difference between the mean values of the distributions corresponds to the distortion of the signal reconstruction. Furthermore, the figure reveals that the signal reconstruction can cause errors, when an information transfer from the TN to the RN takes place. However, when the system parameters are appropriately set, the signal can be reconstructed with a small distortion as observed in Fig. \ref{Distsx} (b).



%

\begin{figure}[!t]
	\centering
	\scalebox{0.45}{\includegraphics{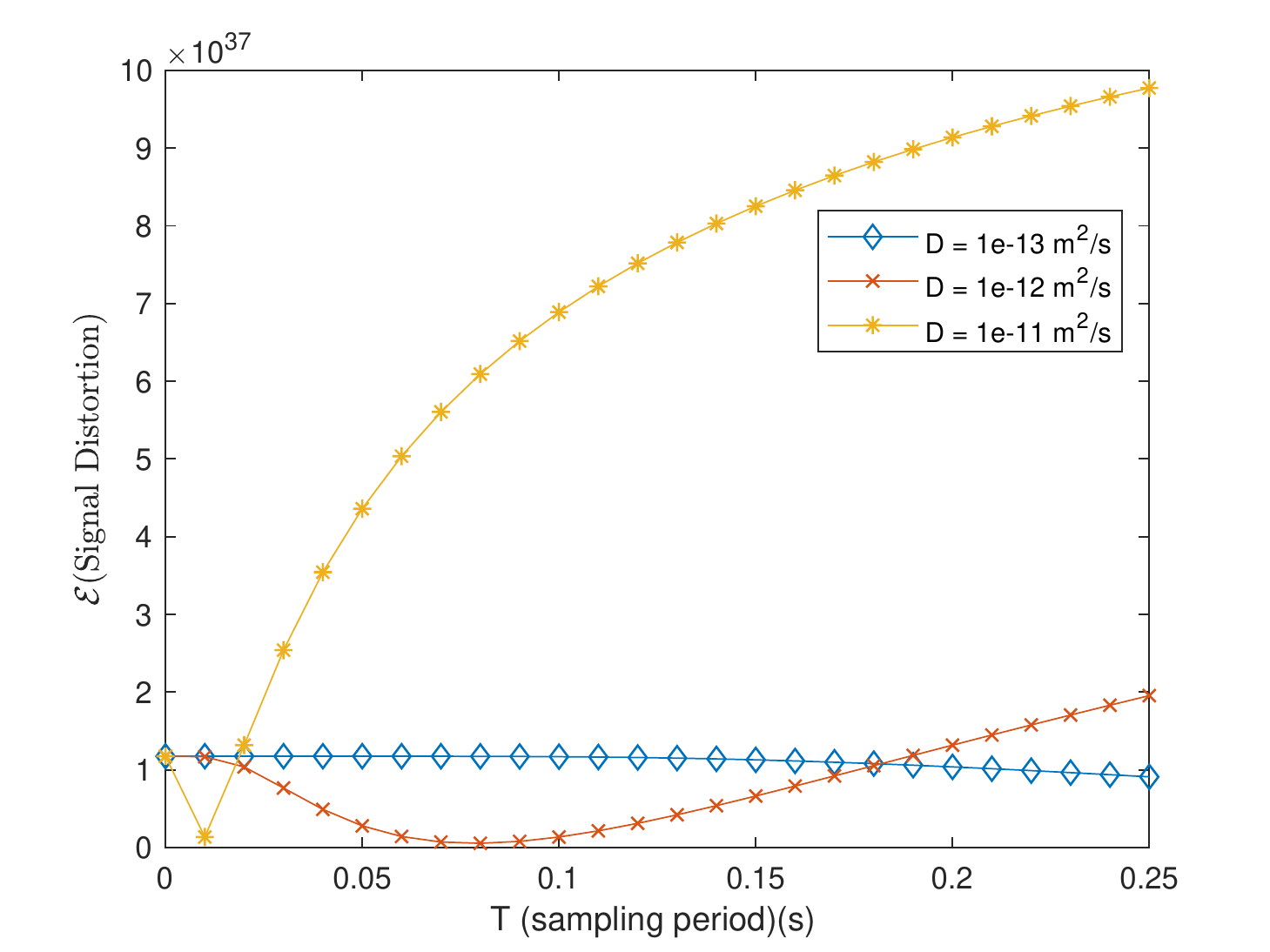}}
	\caption{Signal distortion vs. sampling period for varying diffusion coefficients with $a = 1 \mu m$, $b = 2 \mu m$ and $\rho_{sx}=0.75$.}
	\label{MSE-D}
\end{figure}

In Fig. \ref{MSE-D}, the effect of the diffusion coefficient on the signal distortion is illustrated. For a larger diffusion coefficient, $F(T)$ increases more rapidly. This rapid increase causes the signal distortion with the larger $ D $ to reach its minimum value more quickly. If the RN continues to count the molecules after the minimum of the signal distortion function, the concentration difference grows, since the captured molecules increase and the remaining molecules in $V_R$ decrease. After the minimum point of the signal distortion function, the signal distortion goes up as fast as the magnitude of the diffusion coefficient , due to the faster movement of the molecules in a less dense environment. To obtain minimum signal distortion, the optimization of the RN design parameters is essential, as discussed in the next section.
\section{Receiver Nanomachine Design}
\label{RN Design}
In this section, optimum parameters such as RN radius ($a$), sampling period ($T$) and sampling frequency ($f$) are examined for the RN design. First, we focus on the optimum sampling period, i.e., $T_{opt}$. The sampling period needs to be estimated in order to calculate the receiving time and the information rate of the RN. 

By means of the signal distortion function  given in (\ref{MSE_last}), the optimum parameters can be derived. Since the signal distortion function is convex as observed with the numerical results in Section \ref{Simulation}, $T_{opt}$ can be found by solving the equation for $T$ as given by
\begin{equation}
	\label{Der_t_opt}
	\frac{\partial \mathcal{E}(T)}{\partial T} = 0.
\end{equation}

\begin{figure}[!b]
	\centering
	\scalebox{0.45}{\includegraphics{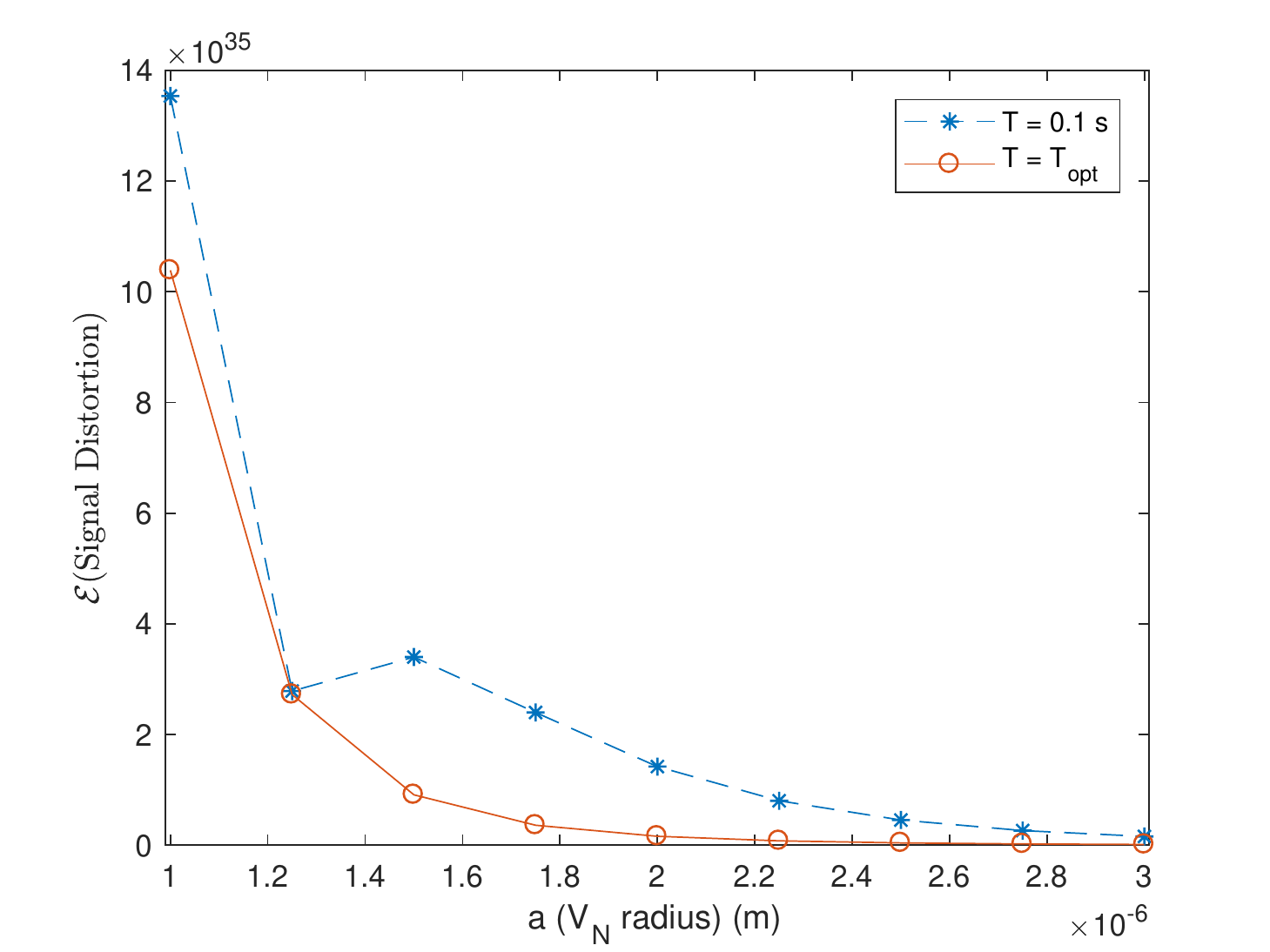}}
	\caption{Constant and optimum sampling period comparison for $ D = 10^{-12} m^2/s $, $b = 2a$, $\mu_s=\sigma_s^2=10^2$ and $  \rho_{sx} = 0.75 $.}
	\label{Opt-T}
\end{figure}

When the sampling period is set as $T_{opt}$, the minimum signal distortion is obtained. The derivative of $ \mathcal{E} $ with respect to $T$ is given in (\ref{Der_MSE_T}) where $z = \frac{b-a}{4\sqrt{DT}}$. However, the derived expression of $T_{opt}$ is a long equation and cannot be written in this paper. Instead, the numerical comparison of the signal distortion for a constant $T$ and $T_{opt}$ is given in Fig. \ref{Opt-T}. To derive $T_{opt}$, an approximation such that $\textrm{erfc}(x) = e^{-c_1x-c_2x^2}$, where $c_1 = 1.09500814703333$ and $c_2 = 0.75651138383854$, is used \cite{tsay2013simple}. Furthermore, it is assumed that $e^{-c_1z-(c_2+1)z^2} \approx e^{-c_1z-(c_2+2)z^2}$, $e^{-2c_1z-2c_2z^2} \approx e^{-2c_1z-(2c_2+2)z^2}$, $e^{-2c_1z-(2c_2+1)z^2} \approx e^{-2c_1z-(2c_2+2)z^2}$ to solve the equation for $T_{opt}$. The numerical results in Fig. \ref{Opt-T} show that the derived $T_{opt}$ gives better results and validate the sampling period is optimum, when it is compared with a constant sampling period. 
\begin{figure*}[!t]
	\normalsize
	\begin{multline}
		\dfrac{\partial \mathcal{E}(T)}{\partial T} = - \dfrac{9(b-a)D \mu_s^2 e^{-z^2}}{16a^2(a+b)(b^3-a^3)\pi^{5/2}(DT)^{3/2}} \\ + \dfrac{9(b^2-a^2)D\mu_s e^{-z^2}+36a(b-a)D\mu_s^2 e^{-z^2}\textrm{erfc}(z)+36a(b-a)D\sigma_s^2 e^{-z^2}\textrm{erfc}(z)}{32(a+b)^2a^5 \pi^{5/2}(DT)^{3/2}} \\	-  \dfrac{18\rho_{sx}\sigma_s^3 a(b-a)De^{-z^2}\textrm{erfc}(z)+9(b^2-a^2)D\mu_s e^{-z^2}+18a(b-a)\sigma_s^2 D \textrm{erfc}(z)}{32a^2 \pi^{5/2} (a+b)(b^3-a^3)(DT)^{3/2}\sqrt{2a\mu_s (a+b) \textrm{erfc}(z) + 4a^2\sigma_s^2\left(\textrm{erfc}(z)\right)^2}}.
		\label{Der_MSE_T}
	\end{multline}
	\vspace*{4pt}
	\hrulefill
\end{figure*}

Another critical parameter of the RN is its radius. It is essential to choose the optimum radius for the minimum MSE signal reconstruction. Similar to $T_{opt}$, the optimum RN radius, i.e., $a_{opt}$, can be found by solving
\begin{equation}
	\label{Der_a_opt}
	\frac{\partial \mathcal{E}(a)}{\partial a} = 0.
\end{equation}
If the second derivative of $ \mathcal{E}(a) $ is positive, its minimum can be calculated. The first derivative of the $ \mathcal{E}(a) $ with respect to $a$ and the solution for $a_{opt}$ are too long equations to write here. Instead, the numerical results regarding the relationship among $a$, $T$, $f$ and the signal distortion are examined.

\begin{figure}[!t]
	\centering
	\scalebox{0.45}{\includegraphics{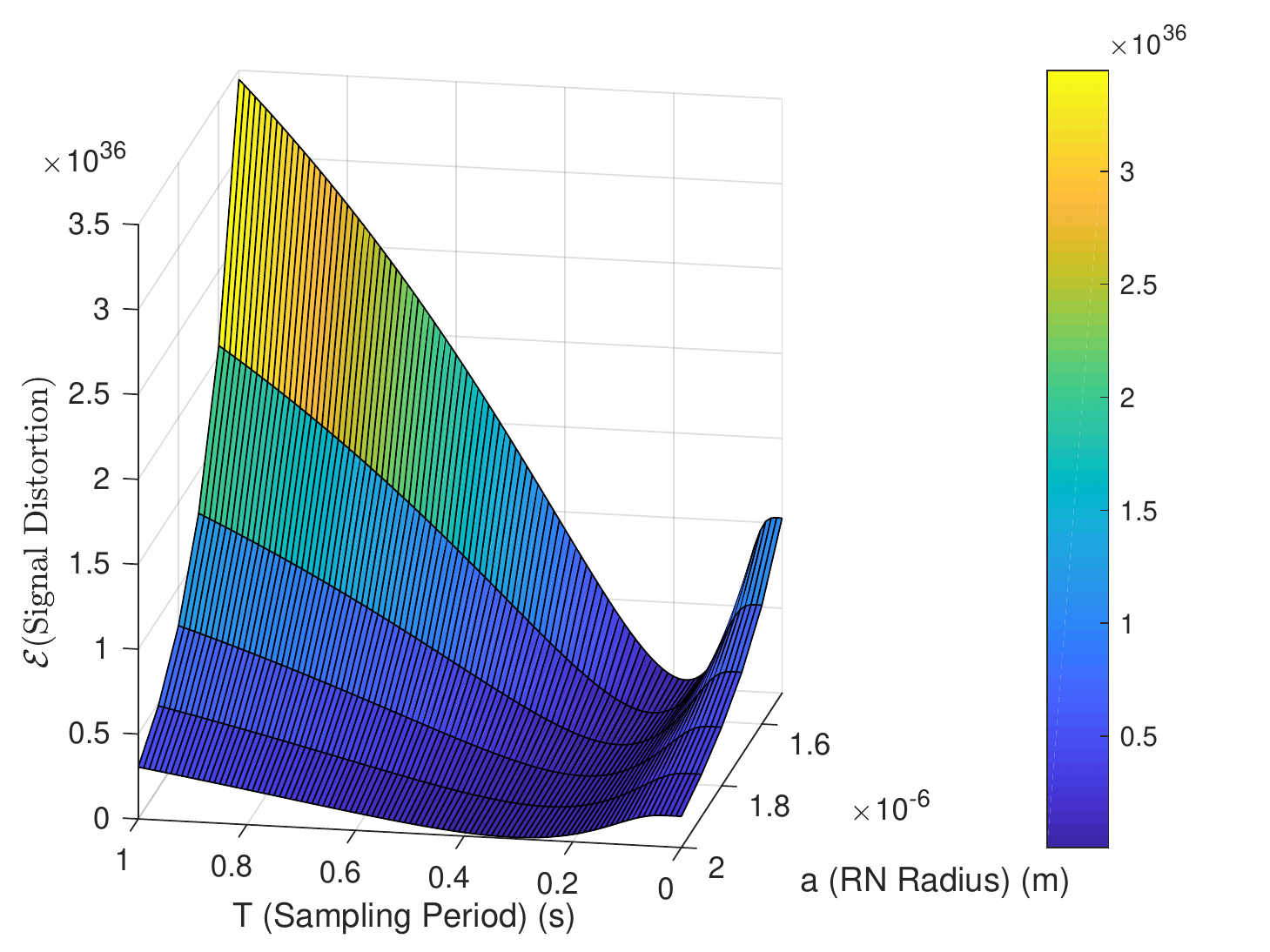}}
	\caption{Signal distortion vs. $a$ and $T$ surface plot  for $\mu_s = 10^2$, $\sigma_s^2 = 10^2$, $ D = 10^{-12} m^2/s $, $b = 2 a$ and $  \rho_{sx} = 0.75 $.}
	\label{MSE-a-T}
\end{figure}

Assuming that $b = 2 a$, the combined effect of the system parameters on the signal distortion is shown in Fig. \ref{MSE-a-T}. As the size of the RN grows, the RN needs more time to collect the sufficient number of molecules required for the desired concentration value around it and thus, to reduce the signal distortion. This stems from the fact that, while the surface of the RN increases proportionally with $a^2$, its volume increases proportionally with $a^3$. To balance this situation, the RN extends its sampling period. Regarding the $T_{opt}$ values, as the RN radius increases, the RN needs more time to reconstruct the signal, but has a lower signal distortion due to the larger capture probability of the molecules. This shows that there is a trade-off among $a$, $T$ and the signal distortion. When a larger RN size is chosen to have a lower signal distortion, a longer optimum sampling period  is required. On the contrary, if the RN is desired to respond faster for reception, its cost is a worse signal reception. Therefore, RN sampling period and its corresponding signal distortion are essential to determine the optimum RN radius, i.e., $a_{opt}$.

\begin{figure}[!h]
	\centering
	\scalebox{0.45}{\includegraphics{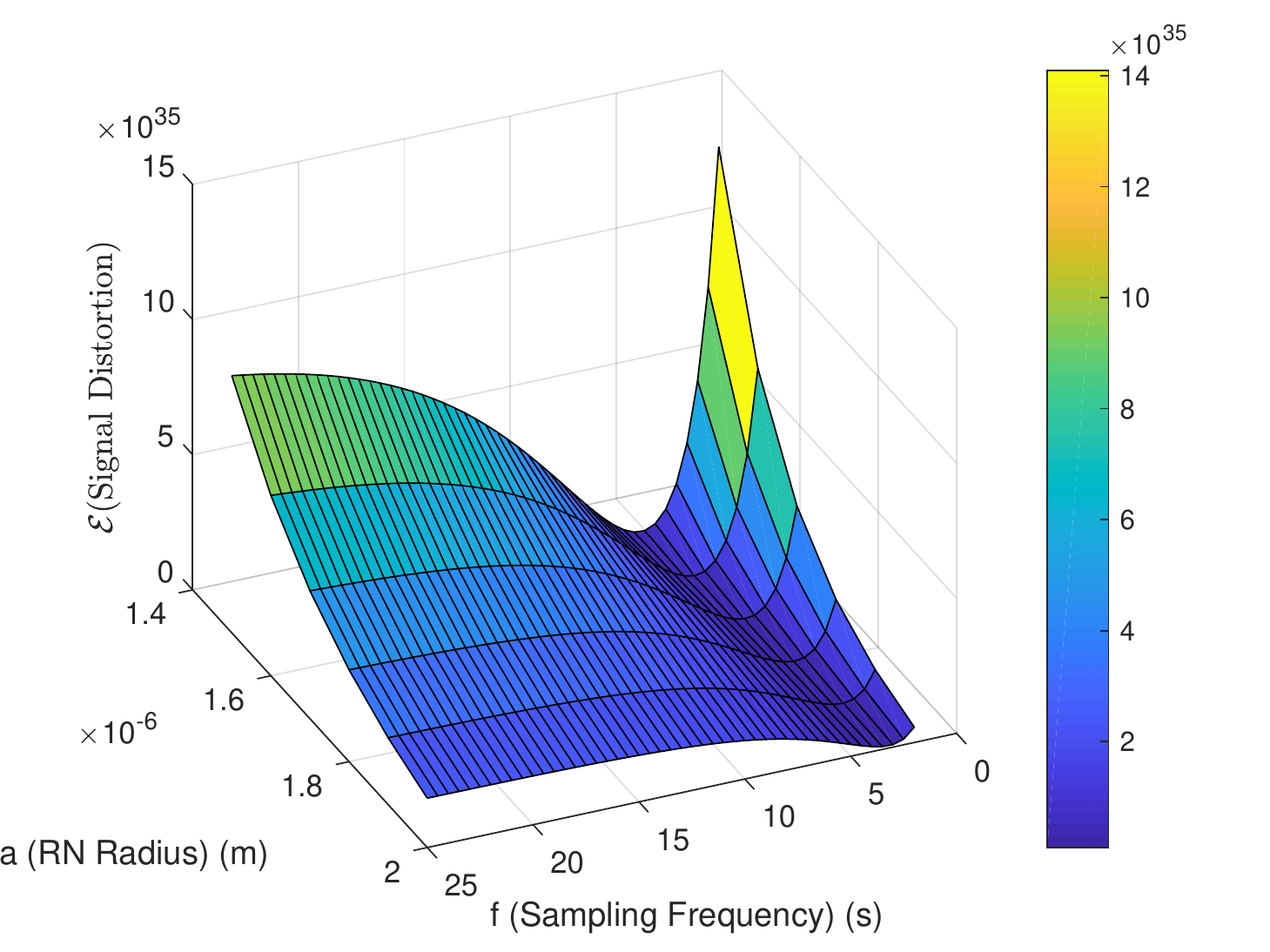}}
	\caption{Signal distortion vs. $a$ and $f$ surface plot  for $\mu_s = 10^2$, $\sigma_s^2 = 10^2$, $ D = 10^{-12} m^2/s $, $b = 2 a$ and $  \rho_{sx} = 0.75 $.}
	\label{MSE-a-f}
\end{figure}

The sampling frequency, which is calculated by $f = \frac{1}{T}$, can be used to design a RN efficiently. The combined effect of the sampling frequency, RN radius and signal distortion can be seen in Fig. \ref{MSE-a-f}. An analysis similar to the sampling period can be made for the sampling frequency. As the RN radius increases, the sampling frequency and signal distortion decrease at the optimum frequency points, i.e., $f_{opt}$. The relation among the signal distortion, sampling frequency and the RN radius is required to be considered for the RN design. When a smaller RN is chosen, the signal distortion and $f_{opt}$ becomes higher. Hence, the cost for a smaller RN with the minimum signal distortion is a lower quality communication and a more complex structure for a faster signal processing. 

It is not always possible to design the RN according to the optimum parameters. In such cases, a signal distortion constraint can be defined. Due to the convexity of the signal distortion function, a range within the minimum and maximum values of the sampling period (or sampling frequency) can be determined for the given signal distortion constraint. Subsequently, the RN radius can be chosen according to this range of the sampling period (or sampling frequency).
\vspace{-0.5cm}
\section{Conclusion}
\label{Conclusion}
In this paper, a new concept about how accurate the molecular signal is sensed and reconstructed by the RNs is proposed. The RN is assumed as a perfect absorbing molecule counting machine and the reconstructed signal is modeled as a counting process. The molecule concentration is treated as a molecular signal and a signal distortion is defined as the mean square error between the existing molecular signal outside the RN and the reconstructed signal. Instead of the deterministic approach, Gaussian random process is used for the molecular signal outside the RN in a more realistic way and DSPP is obtained to model the reconstructed signal.  The derived signal distortion function is validated by means of the  random walk simulations. Numerical results are given to highlight the effect of the system parameters such as the diffusion coefficient, sampling period and RN radius, on the signal distortion. By the minimization of the signal distortion, the optimum RN design parameters are derived. Our analysis about the effect of the signal distortion on the RN design parameters shows that the RN can reconstruct signals with a small distortion, when the RN design parameters are properly configured. 

The signal reconstruction distortion, which is proposed as a novel performance parameter, can be employed to design an efficient MC system from the signal reconstruction perspective. For example, a smaller diffusion coefficient can be chosen for a faster signal reconstruction with minimum signal distortion. Furthermore, this perspective can be utilized to show the trade-off among the RN design parameters such that the optimum sampling period decreases and the minimum signal distortion increases, as the radius of the RN decreases. As the future work, we plan to develop MC methods to transmit and receive information efficiently according to the signal reconstruction of the RNs. Moreover, our future works include the modeling and analysis of the signal reconstruction by the RNs with receptors.

\section*{Acknowledgment}
This work was supported by the Scientific and Technological Research Council of Turkey (TUBITAK) under Grant 115E362.

\bibliography{sr_fg}

\begin{thebibliography}{10}
\providecommand \doibase [0]{http://dx.doi.org/}%

\bibitem{Farsad-2016}
Farsad N, Yilmaz HB, Eckford A, Chae CB, Guo W. A comprehensive survey of
  recent advancements in molecular communication. {\it IEEE Communications
  Surveys \& Tutorials} 2016\string; 18(3)\string: 1887--1919.

\bibitem{Atakan-2014}
Atakan B. {\it Molecular communications and nanonetworks: from nature to
  practical systems}.
\newblock Springer Science \& Business Media .
\newblock 2014.

\bibitem{Akyildiz2008}
Akyildiz IF, Brunetti F, Bl{\'a}zquez C. Nanonetworks: A new communication
  paradigm. {\it Computer Networks} 2008\string; 52(12)\string: 2260--2279.

\bibitem{akyildiz2011nanonetworks}
Akyildiz IF, Jornet JM, Pierobon M. Nanonetworks: A new frontier in
  communications. {\it Communications of the ACM} 2011\string; 54(11)\string:
  84--89.

\bibitem{Hiyama-2005}
Hiyama S, Moritani Y, Suda T, et al. Molecular Communication. In: NSTI. ;
  2005\string: 391-394.

\bibitem{Berg-1977}
Berg HC, Purcell EM. Physics of chemoreception. {\it Biophysical journal}
  1977\string; 20(2)\string: 193--219.

\bibitem{Endres-2008}
Endres RG, Wingreen NS. Accuracy of direct gradient sensing by single cells.
  {\it Proceedings of the National Academy of Sciences} 2008\string;
  105(41)\string: 15749--15754.

\bibitem{Endres-2009}
Endres RG, Wingreen NS. Maximum likelihood and the single receptor. {\it
  Physical review letters} 2009\string; 103(15)\string: 158101.

\bibitem{Aquino-2016}
Aquino G, Wingreen NS, Endres RG. Know the single-receptor sensing limit? Think
  again. {\it Journal of statistical physics} 2016\string; 162\string: 1353.

\bibitem{pierobon2010physical}
Pierobon M, Akyildiz IF. A physical end-to-end model for molecular
  communication in nanonetworks. {\it IEEE Journal on Selected Areas in
  Communications} 2010\string; 28(4)\string: 602 - 611.

\bibitem{pierobon2011noise}
Pierobon M, Akyildiz IF. Noise analysis in ligand-binding reception for
  molecular communication in nanonetworks. {\it IEEE Transactions on Signal
  Processing} 2011\string; 59(9)\string: 4168--4182.

\bibitem{pierobon2011diffusion}
Pierobon M, Akyildiz IF. Diffusion-based noise analysis for molecular
  communication in nanonetworks. {\it IEEE Transactions on Signal Processing}
  2011\string; 59(6)\string: 2532--2547.

\bibitem{lin2018mutual}
Lin L, Wu Q, Liu F, Yan H. Mutual Information and Maximum Achievable Rate for
  Mobile Molecular Communication Systems. {\it IEEE transactions on
  nanobioscience} 2018\string; 17(4)\string: 507--517.

\bibitem{mosayebi2018type}
Mosayebi R, Gohari A, Mirmohseni M, Nasiri-Kenari M. Type-Based Sign Modulation
  and Its Application for ISI Mitigation in Molecular Communication. {\it IEEE
  Transactions on Communications} 2018\string; 66(1)\string: 180--193.

\bibitem{kilinc2013receiver}
Kilinc D, Akan OB. Receiver design for molecular communication. {\it IEEE
  Journal on Selected Areas in Communications} 2013\string; 31(12)\string:
  705--714.

\bibitem{srinivas2012molecular}
Srinivas K, Eckford AW, Adve RS. Molecular communication in fluid media: The
  additive inverse gaussian noise channel. {\it IEEE Transactions on
  Information Theory} 2012\string; 58(7)\string: 4678--4692.

\bibitem{noel2014optimal}
Noel A, Cheung KC, Schober R. Optimal receiver design for diffusive molecular
  communication with flow and additive noise. {\it IEEE transactions on
  nanobioscience} 2014\string; 13(3)\string: 350--362.

\bibitem{mustam2017multilayer}
Mustam SM, Syed~Yusof SK, Nejatian S. Multilayer diffusion-based molecular
  communication. {\it Transactions on Emerging Telecommunications Technologies}
  2017\string; 28(1)\string: e2935.

\bibitem{aghababaiyan2018axonal}
Aghababaiyan K, Shah-Mansouri V, Maham B. Axonal Channel Capacity in
  Neuro-Spike Communication. {\it IEEE transactions on nanobioscience}
  2018\string; 17(1)\string: 78--87.

\bibitem{maham2018neuro}
Maham B, Kizilirmak RC. Neuro-Spike Communications With Multiple Synapses Under
  Inter-Neuron Interference. {\it IEEE Access} 2018\string; 6\string:
  39962--39968.

\bibitem{bossert1963analysis}
Bossert WH, Wilson EO. The analysis of olfactory communication among animals.
  {\it Journal of theoretical biology} 1963\string; 5(3)\string: 443--469.

\bibitem{atakan2010deterministic}
Atakan B, Akan OB. Deterministic capacity of information flow in molecular
  nanonetworks. {\it Nano Communication Networks} 2010\string; 1(1)\string:
  31--42.

\bibitem{ziff2009capture}
Ziff RM, Majumdar SN, Comtet A. Capture of particles undergoing discrete random
  walks. {\it The Journal of chemical physics} 2009\string; 130(20)\string:
  204104.

\bibitem{bai2016throughput}
Bai C, Leeson M, Higgins MD, Lu Y. Throughput and energy efficiency-based
  packet size optimisation of ARQ protocols in bacterial quorum communications.
  {\it Transactions on Emerging Telecommunications Technologies} 2016\string;
  27(8)\string: 1128--1143.

\bibitem{yilmaz2014Threed}
Yilmaz HB, Heren AC, Tugcu T, Chae CB. Three-Dimensional channel
  characteristics for molecular communications with an absorbing receiver. {\it
  IEEE Communications Letters} 2014\string; 18(6)\string: 929 - 932.

\bibitem{cox1955some}
Cox DR. Some statistical methods connected with series of events. {\it Journal
  of the Royal Statistical Society. Series B (Methodological)} 1955\string:
  129--164.

\bibitem{Valderrama-1995}
Valderrama MJ, Jimenez F, Gutierrez R, Martinez-Almecija A. Estimation and
  filtering on a doubly stochastic Poisson process. {\it Applied Stochastic
  Models in Business and Industry} 1995\string; 11(1)\string: 13--24.

\bibitem{berg1993random}
Berg HC. {\it Random walks in biology}.
\newblock Princeton University Press .
\newblock 1993.

\bibitem{tsay2013simple}
Tsay WJ, Huang CJ, Fu TT, Ho IL. A simple closed-form approximation for the
  cumulative distribution function of the composite error of stochastic
  frontier models. {\it Journal of Productivity Analysis} 2013\string;
  39(3)\string: 259--269.

\end{thebibliography}


%

\end{document}